%% file: paper.tex
\documentclass{juliacon}

\usepackage{subcaption}
\usepackage{soul}
\usepackage{amssymb}
\usepackage{graphicx}

\newcommand{\jlpkg}[1]{\texttt{#1}}
\newcommand{\extraejl}{\jlpkg{Extrae.jl}}

\graphicspath{{assets/}}

\begin{document}

\input{header}

\maketitle

\begin{abstract}
The Julia programming language has gained acceptance within the High-Performance Computing (HPC) community due to its ability to tackle two-language problem: Julia code feels as high-level as Python but allows developers to tune it to C-level performance.
But to squeeze every drop of performance, Julia needs to integrate with advanced performance analysis tools, also known as profilers.
In this work, we present \extraejl, a Julia package to interface with the Extrae profiler. 
\end{abstract}

\section{Introduction}

Scientific computing applications require code that is both high-level, for expressing complex mathematical ideas, and high-performance, as it usually involves computationally intensive tasks.
These requirements derive into a problem known as "the two-language problem"; i.e. programming languages usually focus only on one of these two topics, forcing package authors to use two programming languages: one higher-level for the abstraction and one lower-level for performance.
This is characteristic of programming languages such as Python, whose package authors derive to C, C++, or Rust for improving the performance.

In recent years, the Julia programming language~\cite{bezanson2017julia} has steadily gained adoption within the High-Performance Computing (HPC) community due to its ability to tackle the aforementioned two-language problem: Julia code feels as high-level as Python but allows developers to tune it to C-level performance.
Additionally, it supports parallel programming models such as MPI, master-worker distributed computing, and GPU computing, and offers a rich interactive REPL experience, facilitating ease of use and fast prototyping for developers.

While being able to write code so that it can exploit computing resources is key, it is not the full story.
In pursuit of higher computational power, hardware vendors have made more parts of their systems available to software.
This in turn makes software development more relevant than ever for fully exploiting the performance capabilities of current computers.
Driving performance optimization are performance analysis tools that are commonly named as \textit{profilers}.
Profilers measure where a program spends its time or which functions consume the most resources.
They help identify bottlenecks, optimize code, and improve overall application performance.

One of the most well-known profilers in the field is Extrae.
Recognized for its low overhead and large support of programming models and platforms, it is a veteran in the performance analysis field.
Its programming model allows for deep customization and tuning, but unfortunately, it also imposes a high barrier for newcomers.
In this work, we present \extraejl: The Julia bindings to the Extrae profiler, with which we try to ease the barrier for Julia users.

This work is organized as follows.
In Section~\ref{sec:background}, we introduce the current state of performance analysis tools in the HPC world to the reader.
In Section~\ref{sec:design}, we describe Extrae's design and how we have mapped it Julia's concepts and parallelism models.
In Section~\ref{sec:evaluation}, we present some examples of \extraejl running on top of Julia HPC apps.
Finally, in Section~\ref{sec:summary}, we summarize our work and point to possible directions of improvement in future works.

\section{Background}\label{sec:background}

Depending on the target resource of the analysis and the method used to gather the data, different tools are available.
Basically, two methods are used for profiling:

\begin{itemize}
\item Tracing: It involves explicitly calling the profiler at the target regions of the code. Although 100\% is accurate, the overhead is high. Manual instrumentation of the code or binary interception of symbols can be used.
\item Statistical sampling: It entails periodically interrupting the application, collecting data, and resuming. The results are expected to be statistically similar to the ones provided by tracing but with a much lower overhead.
\end{itemize}

There are a number of commercial and open source profiling tools that can be used to analyze performance in HPC based on these two methods.
The following is a list of profilers that are known to work with Julia:

\subsubsection*{Julia standard library \jlpkg{Profiler.jl}}
\label{sec:profiler-jl}

Julia comes with a statistical profiler called \jlpkg{Profiler.jl}, which can be used to periodically sample the backtrace of each currently running task, capturing the name and file-line information of all Julia functions in the call stack.
\jlpkg{Profiler.jl} can also optionally capture information about Julia's core and, optionally, external shared libraries, written in languages such as C, Fortran, etc.
The default output format produced by \jlpkg{Profiler.jl} is text-based, but external tools can be used to produce various types of visualization.
Some of these tools are \jlpkg{FlameGraphs.jl}, VS Code, \jlpkg{ProfileView.jl}, \jlpkg{ProfileVega.jl}, \jlpkg{StatProfilerHTML.jl}, \jlpkg{ProfileSVG.jl}, \jlpkg{PProf.jl}, and \jlpkg{ProfileCanvas.jl}.
It is also possible to profile memory allocations within Julia code with this standard library.
Although \jlpkg{Profiler.jl} does not have native support for distributed computing, it is possible to collect separate traces for different workers.

\subsubsection*{Linux Perf/OProfile}
\label{sec:linux-perf-oprofile}

Julia has support for analyzing programs using popular open source system profilers for the Linux operating system OProfile, a statistical sampler, and its more modern replacement Perf.
By setting the environment variable \jlpkg{ENABLE\_JITPROFILING=1} Julia will create and register an event listener to profile just-in-time (JIT) compiled functions.
Furthermore, the package \jlpkg{LinuxPerf.jl}~\cite{linuxperfjl} allows getting more fine-grained information about hardware counters using Perf for specified function calls.

\subsubsection*{Nvidia Nsight System}
\label{sec:nvidia-nsight-system}

The Nvidia Nsight System is a proprietary statistical sampling profiler with tracing features, which supports MPI and CUDA applications.
As such, it can be used to profile also Julia programs, including those distributed using the MPI protocol and/or employing a Nvidia GPU, but JIT functions will have their names mangled by the compiler.
The package \jlpkg{NVTX.jl}~\cite{nvtxjl} allows you to instrument the code and annotate specific functions or code regions, optionally also adding a payload (e.g. the value of a local variable) which can be displayed in the Nsight System GUI.
The in-development version of Julia can optionally be compiled to automatically instrument with NVTX internal components, such as the compiler and the garbage collector.

\subsubsection*{Intel VTune}
\label{sec:intel-vtune}

Intel VTune is a proprietary sampling profiler with support for MPI and OpenMP applications.
When the environment variable \texttt{ENABLE\_JITPROFILING=1} is set, Julia JIT functions will be recorded with their correct name on VTune traces.
You can instrument Julia code with the package \jlpkg{IntelITT.jl}~\cite{intelittjl} to get richer information in the traces.

\subsubsection*{LIKWID}
\label{sec:likwid}

LIKWID is an open-source suite of tools for performance analysis of HPC applications on Linux, using hardware performance counters, rather than sampling or tracing profiling.
It supports MPI, OpenMP, and GPU offloading. The package \jlpkg{LIKWID.jl}~\cite{likwidjl} interfaces with the shared library part of the LIKWID suite to instrument the code and collect important hardware metrics to measure the performance of an application.

\subsubsection*{Score-P}
\label{sec:score-p}

Score-P is an open-source suite for profiling and event tracing of HPC applications.
You can use the package \jlpkg{ScoreP.jl}~\cite{scorepjl} to instrument Julia code and analyze it with ScoreP.

\subsubsection*{\texttt{MPITape.jl}}
\label{sec:mpitape-jl}

\texttt{MPITape.jl}~\cite{mpitapejl} is an experimental open-source Julia package which overdubs MPI operations performed through the package \jlpkg{MPI.jl}~\cite{byrne2021mpi}.
This allows you to track communication patterns and timings within a distributed application.

\subsubsection*{Tracy}
\label{sec:tracy}

Tracy is an open-source frame and sampling profiler, originally designed for gaming applications but can be used for any other kind of program.
Julia can optionally be built to instrument internal components, such as the garbage collector and the compiler, so that they can be profiled by Tracy.


\section{Design}\label{sec:design}

Extrae generates Paraver traces, which follows two orthogonal object models:

\begin{itemize}
    \item The \textbf{process model} represents the abstract virtual resources used by the most common programming models. Its taxonomy is formed by the WORKLOAD, APPLICATION, TASK, and THREAD objects.
    \item The \textbf{resource model} represents the physical resources where the program is finally executed. Its taxonomy is formed by SYSTEM, NODE and CPU objects.
\end{itemize}

This separation between virtual and physical resources is key for mapping different parallel programming models to a common model.
For example, MPI ranks are assigned to TASK objects, and OpenMP threads are mapped to THREAD objects.
The representation of an application with the OpenMP+MPI programming model composes just as one MPI rank (TASK object) contains several OpenMP threads (THREAD objects).
Note that, thanks to the aforementioned separation, threads can migrate between cores without invalidating the mapping.
Furthermore, new programming models not considered during the development of Extrae/Paraver, such as GPUs, may also map to the Paraver process and resource models.

In this sense, Extrae provides out-of-the-box instrumentation for MPI, OpenMP, OpenACC, PThreads, CUDA, OpenCL, Linux syscalls, Libc's memory-related functions (i.e. \texttt{malloc} and \texttt{free}) and IO functions.
Some of these models require some form of dependency injection mechanism, for which Extrae supports two methods:
\begin{itemize}
    \item Library symbol interception using \texttt{LD\_PRELOAD} 
    \item Binary rewrite instrumentation using DynInst
\end{itemize}

Extrae automatically detects the corresponding resource identifiers. For the process model, the identifiers are taken directly from the underlying API offered by the programming models, but Extrae allows the user to customize the TASK and THREAD identification functions.
This is beneficial for custom programming models built on top of other programming models, as is the case of COMPSs~\cite{badia2015comp} which maps workers to TASK objects.
We provide access to them through the \texttt{set\_threadid\_function!}, \texttt{set\_taskid\_function!} and derivatives.

Additionally, Extrae provides a statistical call stack and hardware counter sampler.
Hardware counters are accessed through the PAPI~\cite{browne2000portable} interface and can be configured to either sample periodically on time or based on accumulative event counters (e.g., sample every 1,000 dispatched instructions).
Jitter can be configured to avoid sampling aliasing effects.

It generates three types of annotation: states, events, and communications. States refer to regions of time where the application spends time on user code, on external code, or idle. Extrae provides a routine for annotating the beginning and end of the user-coded region. For the commodity of users, we made it available through a \texttt{@user\_function} macro. An example of its usage is shown in Listing~\ref{code:user-region-axpy}.

\begin{lstlisting}[language=Julia, label=code:user-region-axpy, caption={Usage example of \texttt{@user\_function} in a benchmark for the \texttt{axpy!} routine.}]
using Extrae: Extrae, user_function

Extrae.init()

function axpy!(a, x, y)
   @user_function @simd for i in eachindex(x, y)
       @inbounds y[i] = muladd(a, x[i], y[i])
   end
   return y
end

for T in (Float16, Float32, Float64)
   benchmark(axpy!, T, "julia")
end

Extrae.finish()
\end{lstlisting}

An omnipresent type of annotation is events. Events are punctual annotations of 2-tuple integer values on a specific processor and time location.
Many automatically collected annotations are events; e.g. hardware counters.
\texttt{Extrae.emit} allows the user to emit custom events, where the first argument is the event type or code, and the second value is the event value.
Event types and values can be given some string descriptions with \texttt{Extrae.register}.
Listing~\ref{code:register-event} shows an example of how to use both the \texttt{emit} and \texttt{register} functions.

\begin{lstlisting}[language=Julia, label=code:register-event, caption={Usage example of event registration and emition.}]
# on setup
const CODE_VEC_LEN::UInt32 = 84210
Extrae.register(CODE_VEC_LEN, "Vector length")

# on workload
Extrae.emit(CODE_VEC_LEN, UInt64(N))
\end{lstlisting}

Finally, the last type of annotation is communication. One communication mark represents a message sent between two processes during execution. Custom emission of communication marks is part of the extended API, for which the support is experimental, but MPI instrumentation already uses it automatically. An example can be shown in Figure~\ref{fig:timeline:zoom}, where yellow lines represent MPI messages.

\subsection{Mapping Julia's programming models to Paraver's process model}

In the case of \jlpkg{MPI.jl}~\cite{byrne2021mpi}, MPI calls are automatically instrumented if Extrae is loaded before the MPI library, but loading it before Julia can lead to some errors due to library version incompatibility issues.
The solution is then to use the \texttt{preloads} setting of \jlpkg{MPI.jl} preferences (using \jlpkg{MPIPreferences.jl}). This will load Extrae after Julia but before the MPI library.

In the case of \jlpkg{Distributed.jl}, workers can map to TASK objects. If MPI is used as the communication layer between distributed workers using the \jlpkg{MPIClusterManagers.jl} package, then the mapping is performed automatically using the procedure described above.
In the case of other cluster managers, the user needs to manually remap the \texttt{taskid} and \texttt{numtasks} routines for Extrae. To simplify this process, we have added a helper method to \texttt{init} so the user only needs to do something similar to Listing~\ref{code:distributed-extrae-init}.

\begin{lstlisting}[language=Julia, label=code:distributed-extrae-init, caption=Manual initialization of Extrae on a Distributed.jl workflow.]
using Distributed

addprocs(4)

@everywhere using Extrae
@everywhere Extrae.init(Val(:Distributed))
\end{lstlisting}

In the case of Julia shared-memory programming model, threads map to THREAD objects automatically since they use PThreads underneath.
However, Julia tasks (i.e. coroutines) are impossible to map, as they add another level of virtual indirection because tasks can migrate between threads when they yield execution.
In this sense, a Julia task id event could be emitted if Extrae could hook to Julia's BPF points but it is something that is not yet supported.
Meanwhile, if task tracing is critical to the user, we propose using manual instrumentation, as the template code in Listing~\ref{code:instrument-tasks}.

\begin{lstlisting}[language=Julia, label=code:instrument-tasks, caption={Template code for instrumenting Julia tasks.}]
taskid() = objectid(current_task())

@spawn begin
    Extrae.emit(TASKID_EVENT, taskid())
    # workload
    yield()

    EXTRAE.emit(TASKID_EVENT, taskid())
    # other workload
    Extrae.emit(TASKID_EVENT, 0)
end
\end{lstlisting}

Finally, thanks to the Yggdrasil repository of binary artifacts, users can fetch the latest Extrae library based on their current platform, MPI ABI and CUDA version.
As an interesting fact, this was the first Yggdrasil recipe to combine both CUDA and MPI platform augmentation, further exemplifying the success of Julia's artifact system and BinaryBuilder.

For a deeper insight into Extrae and Paraver, we point the reader to \cite{extrae_documentation,paraver_reference,paraver_seminar}.

\section{Evaluation}\label{sec:evaluation}

With the aim of testing the integration of \extraejl with Julia, we evaluate the performance of a Taylor-Green vortex simulation parallelized with MPI using \jlpkg{Trixi.jl}~\cite{ranocha2022adaptive,schlottkelakemper2020trixi} and \jlpkg{OrdinaryDiffEq.jl}~\cite{DifferentialEquations.jl-2017}.
The source code can be found in \url{https://github.com/trixi-framework/performance-2024-trixi_taylor-green_vortex}.
The application was run on the Marenostrum 5 supercomputer and the analysis was performed with the Paraver trace viewer~\cite{paraver_reference,pillet1995paraver}.

\begin{figure}[h]
    \centering
    \includegraphics[width=\linewidth]{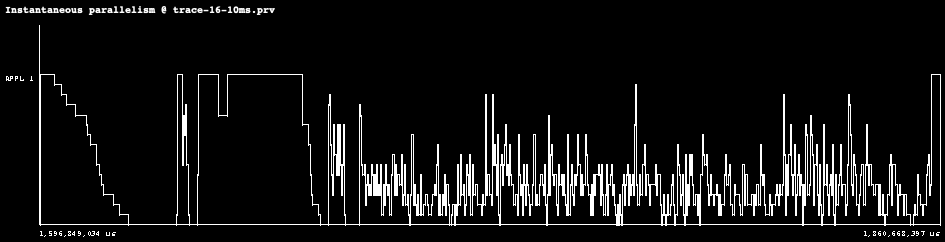}
    \caption{Instantaneous parallelism during the workload of the application measured as the number of MPI ranks not being idle at the given moment. Maximum value is 16 and minimum is 1.}
    \label{fig:parallelism}
\end{figure}

First, Paraver is able to compute some general metrics about the run. For example, in Figure~\ref{fig:parallelism} the evolution of instantaneous parallelism is shown, which reveals parallelization problems during the main workload.

\begin{figure}[h]
    \centering
    \begin{subfigure}{\linewidth}
        \includegraphics[width=\linewidth]{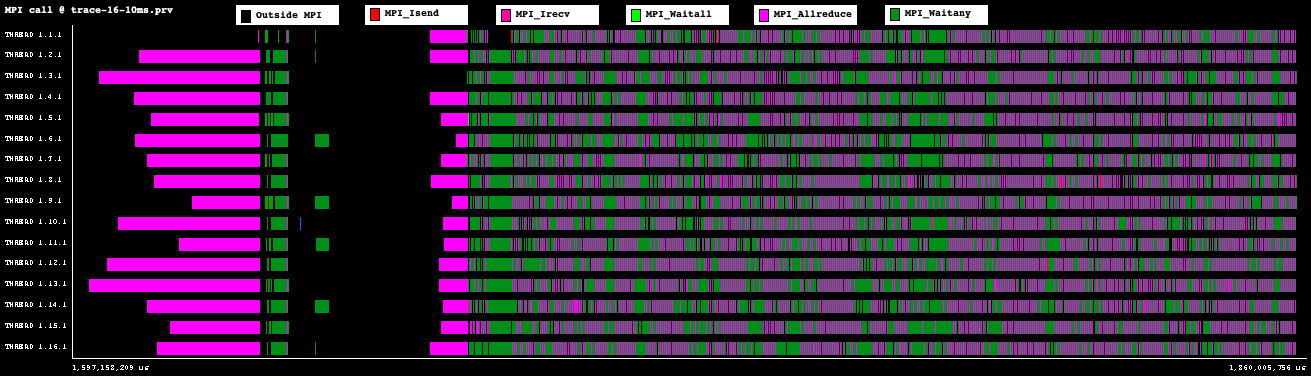}
        \caption{Original}
        \label{fig:timeline:orig}
    \end{subfigure}
    \\\vspace{1em}
    \begin{subfigure}{\linewidth}
        \includegraphics[width=\linewidth]{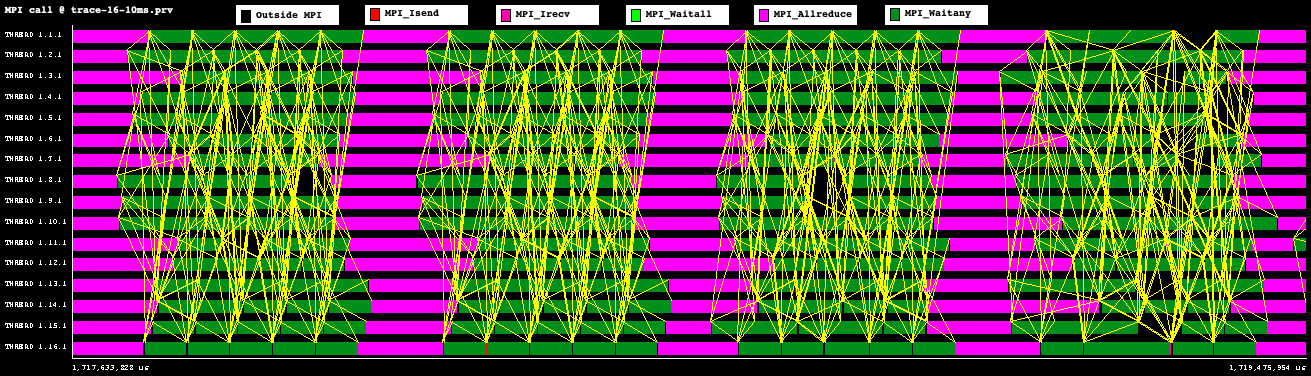}
        \caption{Zoom}
        \label{fig:timeline:zoom}
    \end{subfigure}
    \\\vspace{1em}
    \caption{Timeline of MPI calls for the (a) runtime of the application and (b) zoom on 
 a couple of iterations of the main workload. Horizontal axis is time and each row shows the events that took place on each MPI rank. Blocks represent MPI calls where color maps to MPI routines and the start and end positions fit the duration of the call. Yellow lines represent messages sent betweens ranks.}
    \label{fig:timeline}
\end{figure}

For a more detailed and finer view of the execution, Paraver can plot a timeline of MPI calls per rank, as shown in Figure~\ref{fig:timeline:orig}. By sight, we can correlate the region of loss of parallelization with the region where there is a high density of \texttt{MPI\_Waitany} and \texttt{MPI\_Allreduce} calls. Zooming in on that region, shown in Figure~\ref{fig:timeline:zoom}, we can confirm that execution is dominated by these two MPI routines.

\begin{figure}[h]
    \centering
    \includegraphics[width=.8\linewidth]{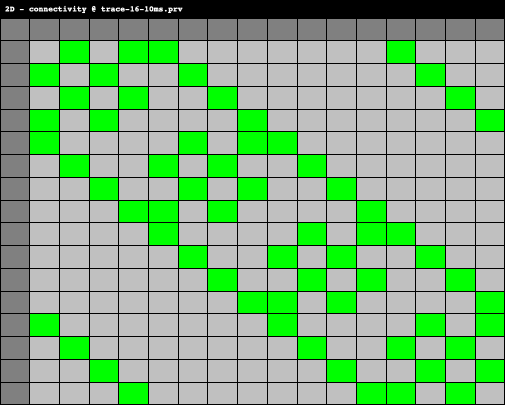}
    \caption{Connectivity pattern between MPI ranks measured as the number of messages sent from MPI rank \textit{x} to rank \textit{y}. Grey cells mean no communication while green cells mean represent a value of 2,016 messages sent.}
    \label{fig:connectivity}
\end{figure}

If we take the connectivity pattern between MPI ranks, shown in Figure~\ref{fig:connectivity}, we can observe no communication imbalance.

\begin{figure}[h]
    \centering
    \newlength{\svgwidth}
    \setlength{\svgwidth}{\linewidth}
    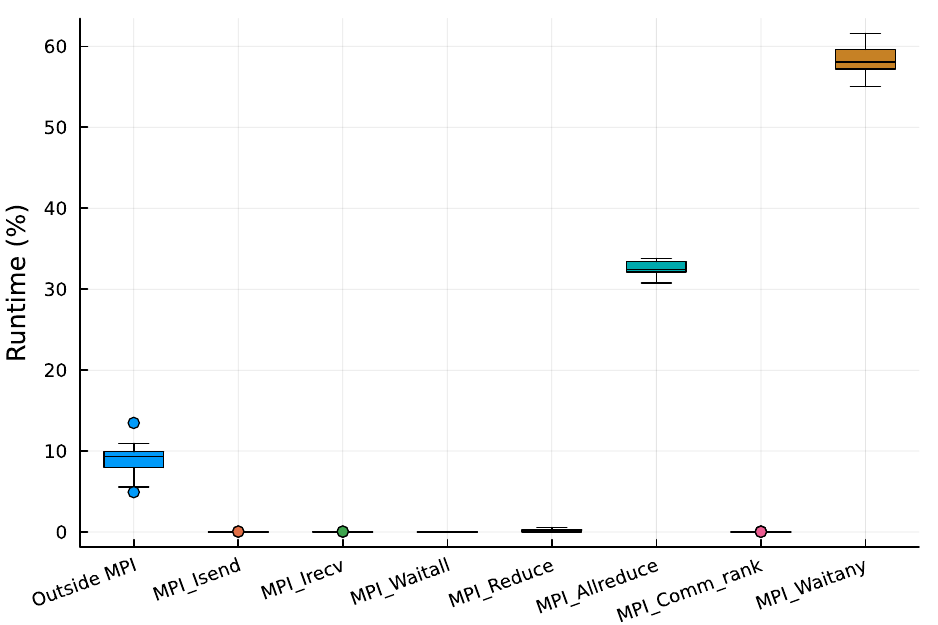
    \let\svgwidth\undefined
    \caption{Fraction of time spent on each MPI routine. Dispersion comes from the time spent by different MPI ranks.}
    \label{fig:mpi-runtime-perc}
\end{figure}

Paraver also allows for external post-processing of the collected data.
It is clear from Figure~\ref{fig:mpi-runtime-perc} that the bottleneck is \texttt{MPI\_Waitany} (around 60\% of the total time), followed by \texttt{MPI\_Allreduce} (around 30\% of the total time).
Furthermore, the variability is small enough compared to the amount of time to discard load imbalance problems.

\begin{figure}
    \centering
    \includegraphics[width=\linewidth]{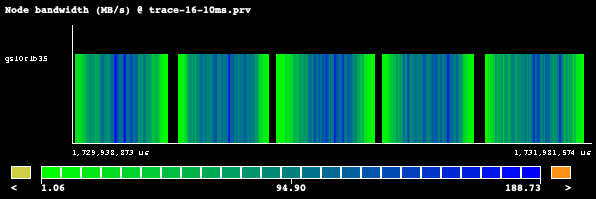}
    \caption{Node network bandwidth in the time region of Figure~\ref{fig:timeline:zoom} measured in MB/s. Only one node is shown because all MPI ranks run on the same node.}
    \label{fig:node-bw}
\end{figure}

Finally, Paraver can also estimate the node bandwidth by taking the communication annotations, as shown in Figure~\ref{fig:node-bw}.
In correlation with Figure~\ref{fig:timeline:zoom}, we can see that the bandwidth increases during \texttt{MPI\_Waitany} calls and decreases and even stops communication during \texttt{MPI\_Allreduce} calls. In addition, the bandwidth peaks at around 188.73 MB/s, far away from the theoretical peak bandwidth of 12.5~GB/s (100~Gb/s).

\section{Summary}\label{sec:summary}

In this work, we have presented \extraejl, a Julia package to interface with the Extrae profiler.
Furthermore, we have demonstrated the integration through the use of Extrae to evaluate the performance of a Taylor-Green vortex simulation using MPI, \jlpkg{Trixi.jl} and \jlpkg{OrdinaryDiffEq.jl}.
Thanks to the Yggdrasil binary repository, we provide binary artifacts for the target platform, MPI ABI and CUDA version, so users can use it without requiring an administrator to install it in their cluster.

In the future, we aim to integrate with other BSC's performance modeling tools such as Folding and Dimemas.
Alternatively, there may be a reimplementation of this functionality in Julia through the use of the Paraver parser.
A Terminal User Interface (TUI) is planned for interactive configuration of Extrae.
Moreover, there is an ongoing effort to parse Paraver tracefiles and convert them to Open Trace Format 2 (OTF2) which would ensure compatibility with other trace visualization tools like ScoreP.

\section{Acknowledgments}

MG's work was funded by the UCL ARC "Fostering International Collaborations in Advanced Digital Research" program.
Authors would like to thank Marc Clascà, for help instrumenting \jlpkg{Distributed.jl} and general Extrae advice, and Carsten Bauer, for interest and suggestions on future work directions.

\input{bib.tex}

\end{document}

%% file: header.tex

\title{Extrae.jl: Julia bindings for the Extrae HPC Profiler}

\author[1]{Sergio Sanchez-Ramirez}
\author[2]{Mosè Giordano}
\affil[1]{Barcelona Supercomputing Center (BSC)}
\affil[2]{University College London (UCL)}


\hypersetup{
pdftitle = {My JuliaCon proceeding},
pdfsubject = {JuliaCon 2024 Proceedings},
pdfauthor = {1st author, 2nd author, 3rd author},
pdfkeywords = {Julia, Optimization, Game theory, Compiler},
}

%% file: assets/mpi_call_profile_svg-tex.pdf_tex
\begingroup%
  \makeatletter%
  \providecommand\color[2][]{%
    \errmessage{(Inkscape) Color is used for the text in Inkscape, but the package 'color.sty' is not loaded}%
    \renewcommand\color[2][]{}%
  }%
  \providecommand\transparent[1]{%
    \errmessage{(Inkscape) Transparency is used (non-zero) for the text in Inkscape, but the package 'transparent.sty' is not loaded}%
    \renewcommand\transparent[1]{}%
  }%
  \providecommand\rotatebox[2]{#2}%
  \newcommand*\fsize{\dimexpr\f@size pt\relax}%
  \newcommand*\lineheight[1]{\fontsize{\fsize}{#1\fsize}\selectfont}%
  \ifx\svgwidth\undefined%
    \setlength{\unitlength}{450bp}%
    \ifx\svgscale\undefined%
      \relax%
    \else%
      \setlength{\unitlength}{\unitlength * \real{\svgscale}}%
    \fi%
  \else%
    \setlength{\unitlength}{\svgwidth}%
  \fi%
  \global\let\svgwidth\undefined%
  \global\let\svgscale\undefined%
  \makeatother%
  \begin{picture}(1,0.66666667)%
    \lineheight{1}%
    \setlength\tabcolsep{0pt}%
    \put(0,0){\includegraphics[width=\unitlength,page=1]{mpi_call_profile_svg-tex.pdf}}%
  \end{picture}%
\endgroup%

%% file: bib.tex

\bibliographystyle{juliacon}
\bibliography{ref.bib}